\documentclass[pra,twocolumn]{revtex4}

\usepackage{type1cm}
\usepackage[dvipdfmx]{graphicx}

\usepackage{amsmath,amssymb,bm,amsthm}

\def\beq{\begin{equation}}
\def\eeq{\end{equation}}
\def\nbeq{\begin{equation*}}
\def\neeq{\end{equation*}}

\def\<{\langle}
\def\>{\rangle}

\renewcommand{\d}{\partial}

\begin{document}
\title{Floquet resonant states and validity of the Floquet-Magnus expansion in the periodically driven Friedrichs models}

\author{Takashi Mori}
\email{
mori@spin.phys.s.u-tokyo.ac.jp}
\affiliation{
Department of Physics, Graduate School of Science,
University of Tokyo, Bunkyo-ku, Tokyo 113-0033, Japan
}

\begin{abstract}
The Floquet eigenvalue problem is analyzed for periodically driven Friedrichs models on discrete and continuous space.
In the high-frequency regime, there exists a Floquet bound state consistent with the Floquet-Magnus expansion in the discrete Friedrichs model, while it is not the case in the continuous model.
In the latter case, however, the bound state predicted by the Floquet-Magnus expansion appears as a metastable state whose lifetime diverges in the limit of large frequencies.
We obtain the lifetime by evaluating the imaginary part of the quasi-energy of the Floquet resonant state.
In the low-frequency regime, there is no Floquet bound state and instead the Floquet resonant state with exponentially small imaginary part of the quasi-energy appears, which is understood as the quantum tunneling in the energy space.
\end{abstract}
\maketitle


It is a central issue in nonequilibrium physics to understand the long-time behavior of quantum systems under driving fields periodic in time~\cite{Grifoni_review1998,Haenggi_review1998,Kohler_review2005,Bukov_review2014}.
In the field of condensed matter~\cite{Oka2009,Lindner2011,Kitagawa2011,Tong2013,Perez-Piskunow2014} and cold atomic systems~\cite{Aidelsburger2013,Miyake2013}, it has attracted much attention to engineer and manipulate novel properties of matter by irradiating intense laser fields.
Floquet theory is a useful tool to treat periodically driven quantum systems~\cite{Shirley1965,Sambe1973}.
The time-dependent Hamiltonian is denoted by $H(t)=H_0+V(t)$, where $V(t)=V(t+T)$ stands for the Hamiltonian of the driving field with the period $T=2\pi/\omega$, then the Floquet theorem states that the time-dependent Schr{\"o}dinger equation $id|\psi(t)\>/dt=H(t)|\psi(t)\>$ is solved as $|\psi(t)\>=\sum_{\alpha}C_{\alpha}e^{-i\varepsilon_{\alpha}t}|u_{\alpha}(t)\>$ with $-\omega/2\leq\varepsilon_{\alpha}<\omega/2$, $|u_{\alpha}(t)\rangle=|u_{\alpha}(t+T)\rangle$, and $\<u_{\alpha}(t)|u_{\beta}(t)\>=\delta_{\alpha,\beta}$ for any $t$.
If we define the Floquet operator $F$ as $F=\mathcal{T}\exp\left[-i\int_0^TdtH(t)\right]$, then $F|u_{\alpha}(0)\rangle=e^{-i\varepsilon_{\alpha}T}|u_{\alpha}(0)\>$ and $|u_{\alpha}(t)\>=\mathcal{T}\exp\left[-i\int_0^t dt' (H(t')-\varepsilon_{\alpha})\right]|u_{\alpha}(0)\>$.
Here the symbol $\mathcal{T}$ in front of the exponential is the time-ordering operator.
$|u_{\alpha}(t)\>$ and $\varepsilon_{\alpha}$ are called the Floquet state and the Floquet quasi-energy, respectively.
If there is no degeneracy in the Floquet quasi-energies, the infinite-time average of an observable $O$ is written as $\sum_{\alpha}|C_{\alpha}|^2\<O\>_{\alpha}$, where $\<O\>_{\alpha}=(1/T)\int_0^Tdt\<u_{\alpha}(t)|O|u_{\alpha}(t)\>$.
Thus sufficiently long time behavior of the system is determined by the property of each Floquet state $|u_{\alpha}(t)\>$ and the initial population $\{|C_{\alpha}|^2\}$.

Recently, it has been argued that we can realize some interesting states of matter by irradiating intense external fields rapidly oscillating in time~\cite{Oka2009,Lindner2011,Kitagawa2011}.
These studies rely on the Floquet-Magnus (FM) expansion.
That is, when $V(t)=f(t)V$, where $f(t)=f(t+T)$ is the amplitude and $V$ is the operator of the external field, the Floquet operator in the rotating frame, $F_r=\mathcal{T}\exp\left[-i\int_0^TdtH_r(t)\right]$, where $H_r(t)=U^{\dagger}(t)H_0U(t)$ with $U(t)=\mathcal{T}e^{-i\int_0^tdt'V(t')}$, is expanded as
$F_r=\exp\left[-iH_{\rm eff}T\right]$
with
\beq
H_{\rm eff}=H_{\rm eff}^{(0)}+\frac{1}{\omega}H_{\rm eff}^{(1)}+\frac{1}{\omega^2}H_{\rm eff}^{(2)}+\dots,
\eeq
see Refs.~\cite{Blanes_review2009,Bukov_review2014} for the explicit form of $H_{\rm eff}^{(k)}$.
In the high-frequency regime, it is expected that we can make the approximation $H_{\rm eff}\simeq H_{\rm eff}^{(0)}=(1/T)\int_0^TdtH_r(t)$.
As a result, the time-dependent problem is reduced to the static problem with the effective Hamiltonian $H_{\rm eff}^{(0)}$ for large $\omega$, and this method has been successful in predicting some interesting phenomena.
We call $H_{\rm eff}^{(0)}$ simply the ``FM effective Hamiltonian''.

On the other hand, other recent studies have provided us with evidence that no matter how large $\omega$ is, all the Floquet states in a macroscopic nonintegrable system look the same~\cite{D'Alessio2014,Ponte2015,Lazarides2014b}, which means that the system heats up to infinite temperature regardless of the initial state~\endnote
{In integrable systems, it is not the case, see Refs.~\cite{Russomanno2012,Lazarides2014a}, and there are some numerical reports that for particular driving protocols a certain ergodic system does not heat up to infinite temperature even in the thermodynamic limit~\cite{Prosen1998,DAlessio2013}.}.
This apparently contradicts the argument by the FM expansion which states that a Floquet state is approximately identical to an energy eigenstate of the effective Hamiltonian $H_{\rm eff}^{(0)}$.
The difficulty comes from the divergence of the FM expansion.
The convergence is ensured only for $E_{\rm max}-E_{\rm min}\lesssim\omega$, where $E_{\rm max}$ and $E_{\rm min}$ are the maximum and minimum eigenvalues of $H_0$, respectively~\cite{Blanes_review2009,Bukov_review2014}.
In a macroscopic system, this condition is not satisfied unless $\omega$ is scaled with the system size.
The convergence of the Magnus expansion was investigated in exactly solvable models~\cite{Feldman1984,Fernandez1990}.

It is noted that the divergence of the FM expansion already exists for a one particle problem with an unbounded Hamiltonian.
Hence, investigating concrete one-particle models would be helpful in understanding what happens when the FM expansion fails to converge.

In this work, we consider the discrete and continuous Friedrichs models, the former of which has a bounded energy spectrum and the FM expansion is convergent for large $\omega$, while the latter of which is not bounded and the convergence of the FM expansion is not ensured for any $\omega$.
We will show that exact Floquet eigenstates of the continuous Friedrichs model are actually quite different from eigenstates of $H_{\rm eff}^{(0)}$ even for large $\omega$; there is a bound state in $H_{\rm eff}^{(0)}$ but there is not in the exact Floquet eigenstates. 
However, an eigenstate of $H_{\rm eff}^{(0)}$ can be interpreted as a resonant state with a long but finite lifetime.
The lifetime $\tau_{\rm res}$ of this resonant state roughly behaves as $\tau_{\rm res}\sim\omega^{1/2}$ for large $\omega$ in our model.

\begin{figure}[t]
\begin{center}
\includegraphics[clip,width=6cm]{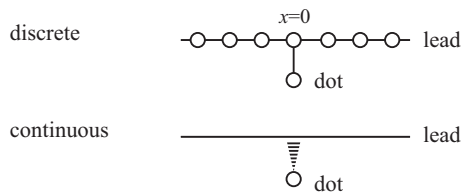}
\caption{The discrete and continuous Friedrichs models.
A particle moves around the lead and dot.}
\label{fig:model}
\end{center}
\end{figure}

The Hamiltonian of the discrete Friedrichs model is given by
\begin{align}
H_d=-g\sum_{x=-\infty}^{\infty}\left(|x+1\>\<x|+|x\>\<x+1|\right) \nonumber \\
-\lambda\left(|d\>\<0|+|0\>\<d|\right)
\label{eq:discrete}
\end{align}
and the Hamiltonian of the continuous Friedrichs model reads
\begin{align}
H_c=\int_{-\infty}^{\infty}dx\left[-g|x\>\d_x^2\<x|+\lambda W(x)\left(|d\>\<x|+|x\>\<d|\right)\right],
\label{eq:continuous}
\end{align}
where $|x\>$ with $x\in\mathbb{Z}$ for the discrete model and $x\in\mathbb{R}$ for the continuous model is the state of the particle at $x$ in the ``lead'' and $|d\>$ is the state of the particle at the ``dot'', see Fig.~\ref{fig:model}.
For simplicity, we restrict ourselves  to $W(x)=\delta(x)$.
The energy spectra of $H_d$ and $H_c$ can be exactly obtained.
In $H_d$, there are two bound states with the energies $\pm E_d$ with $E_d=-\sqrt{2g^2+\sqrt{4g^4+\lambda^4}}<0$ and the continuous spectrum on the range $-2g<E<2g$.
In $H_c$ with $W(x)=\delta(x)$, there is a bound state with the energy $E_c=-(\lambda^4/4g)^{1/3}$ and the continuous spectrum extends over $E>0$.
We study the fate of these bound states under the time-periodic driving field $V(t)=f\cos(\omega t)|d\>\<d|$.
Before going on to the analysis of the Floquet eigenvalue problem, let us see what is the prediction of the FM effective Hamiltonian.
In the rotating frame, $|d\>$ is replaced by $e^{i(f/\omega)\sin\omega t}|d\>$.
Hence, the FM effective Hamiltonians are obtained by replacing $\lambda$ by $\lambda J_0(f/\omega)$ in $H_d$ and $H_c$, where $J_0(\cdot)$ is the 0th order Bessel function.
As a result, the FM effective Hamiltonian also has bound states with energies $\pm E_d^{\rm eff}=\mp\sqrt{2g^2+\sqrt{4g^4+[\lambda J_0(f/\omega)]^4}}$ for the discrete model and a bound state with the energy $E_c^{\rm eff}=-\left\{[\lambda J_0(f/\omega)]^4/4g\right\}^{1/3}$ in the continuous model.

Now let us analyze the Floquet eigenvalue problem.
In the discrete model, the Floquet eigenvalues have been examined in Refs.~\cite{Yamada2012,Noba2014} and the formal expression of the Floquet quasi-energies have been obtained.
However, the concrete analytical evaluation was limited to the weak coupling case $\lambda\ll\omega<4g$.
Here we do not assume the weak coupling.
The Floquet eigenvalue problem is given by
\beq
\left[ H(t)-i\frac{\d}{\d t}\right]|u(t)\>=\varepsilon|u(t)\>,
\label{eq:FEP}
\eeq
where $\varepsilon$ is a Floquet quasienergy with $-\omega/2\leq\varepsilon<\omega/2$ and $|u(t)\>$ is the corresponding Floquet state.
Performing the Fourier transformation, we have
\beq
\sum_{m=-\infty}^{\infty}\left(H_{nm}-n\omega\delta_{n,m}\right)|u_m\>=\varepsilon|u_n\>,
\label{eq:FEP_Fourier}
\eeq
where $H_{nm}=(1/T)\int_0^Tdte^{i(n-m)\omega t}H(t)$ and $|u_n\>=(1/T)\int_0^Tdte^{in\omega t}|u(t)\>$.
By introducing the new vector $|\Psi\>\>=\sum_{n=-\infty}^{\infty}|u_n\>\otimes|n\>$ with $\<n|m\>=\delta_{nm}$, Eq.~(\ref{eq:FEP_Fourier}) is rewritten as
\beq
\mathcal{H}|\Psi\>\>=\varepsilon|\Psi\>\>,
\label{eq:FEP2}
\eeq
where $(1\otimes\<n|)\mathcal{H}(1\otimes|m\>)\equiv H_{nm}$.
Thus the original time-dependent problem is reduced to an eigenvalue problem of $\mathcal{H}$, which acts on the extended Hilbert space~\cite{Sambe1973,Grifoni_review1998}.
The Hilbert space spanned by $\{|x\>\}$ and $|d\>$ is called the ``real space'' and that spanned by $\{|n\>\}$ is called the ``energy space''.

In order to solve Eq.~(\ref{eq:FEP2}), we employ the Feshbach formalism~\cite{Feshbach1958,Feshbach1962}.
The projection operators onto the dot and the lead are introduced as $P=|d\>\<d|\otimes1$ and $Q=1-P$, respectively.
Then the eigenvalue problem is transformed to
\beq
\mathcal{H}_{\rm dot}|\psi\>=\varepsilon|\psi\>,
\label{eq:FEP_dot}
\eeq
where
\beq
|d\>\<d|\otimes\mathcal{H}_{\rm dot}=P\mathcal{H}P+P\mathcal{H}Q\frac{1}{\varepsilon-Q\mathcal{H}Q}Q\mathcal{H}P
\label{eq:dot_eff}
\eeq
and $\mathcal{H}_{\rm dot}$ is the effective Hamiltonian for the dot and $|\psi\>$ is a vector in the energy space alone.
Any solution of Eq.~(\ref{eq:FEP_dot}) is one of the Floquet bound states localized near the dot.
Floquet scattering states $|\Psi_s\>\>$ are not obtained by Eq.~(\ref{eq:FEP_dot}) because $\<\<\Psi_s|P|\Psi_s\>\>/\<\<\Psi_s|\Psi_s\>\>=0$.
By a straightforward calculation, we can obtain the explicit form of $\mathcal{H}_{\rm dot}$ as
\begin{align}
\mathcal{H}_{\rm dot}=\sum_{n=-\infty}^{\infty}\left[-n\omega+\lambda^2v(\varepsilon+n\omega)\right]|n\>\<n|
\nonumber \\
+\frac{f}{2}\sum_{n=-\infty}^{\infty}\left(|n+1\>\<n|+|n\>\<n+1|\right),
\label{eq:Hdot}
\end{align}
where in the discrete Friedrichs model
\begin{align}
v(E)&=\frac{-i}{(2g-E)\sqrt{\frac{2g+E}{2g-E}}}{\rm sgn}\left({\rm Im}\sqrt{\frac{2g+E}{2g-E}}\right)
\nonumber \\
&\equiv v_d(E),
\end{align}
and in the continuous Friedrichs model
\beq
v(E)=-\frac{1}{2\sqrt{-gE}}\equiv v_c(E).
\eeq
Here we have defined $\sqrt{z}$ so that ${\rm Re}\sqrt{z}\geq 0$.
Note that $\lim_{\delta\rightarrow +0}v_d(E\pm i\delta)=\mp i/\sqrt{4g^2-E^2}$ for $-2g<E<2g$, and $\lim_{\delta\rightarrow +0}v_c(E+i\delta)=\mp i/(2\sqrt{gE})$ for $E>0$.
Thus the on-site energy becomes complex in the continuous spectrum of $H_d$ or $H_c$.
This complex on-site energy is interpreted as the decay into continuum.
An important property of $v_d(E)$ and $v_c(E)$ is that the bound state energies of the undriven systems satisfy $\lambda^2v_d(E_d)=E_d$ and $\lambda^2v_c(E_c)=E_c$.

Let us consider the high-frequency regime, $\omega\gg g, |E_b|$, where $E_b$ is the bound state energy; $E_b=E_d$ for the discrete model and $E_b=E_c$ for the continuous model. 
In that case $\lambda^2v(\varepsilon+n\omega)$ in Eq.~(\ref{eq:Hdot}) can be regarded as a perturbation.
The eigenstates $\{|\psi_k^{(0)}\>\}$ and the eigenvalues $\varepsilon_k^{(0)}$ of $H_{\rm dot}$ without perturbation are exactly obtained, $|\psi_k^{(0)}\>=\sum_{n=-\infty}^{\infty}J_{n-k}(f/\omega)|n\>$ and $\varepsilon_k^{(0)}=-k\omega$ with $k$ integer, which is known as the Wannier-Stark ladder~\cite{Wannier1960}.
The $m$th order Bessel function has been denoted by $J_m(\cdot)$.
Because the Floquet quasienergy should be in the range $[-\omega/2,\omega/2)$, we are particularly interested in $k=0$.

\begin{figure}[t]
\begin{center}
\includegraphics[clip,width=8cm]{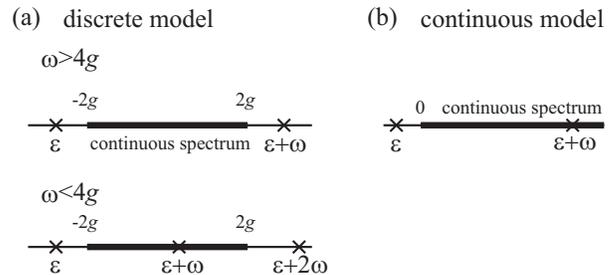}
\caption{(a) In the discrete model, when $\omega$ is sufficiently large, the sequence of $\{\varepsilon+n\omega\}$ can avoid the continuous spectrum, but when $\omega<4g$, it cannot.
(b) In the continuous model, $\varepsilon+n\omega$ for $n>0$ lies on the continuous spectrum no matter how large $\omega$ is.}
\label{fig:hopping}
\end{center}
\end{figure}

We shall evaluate the effect of the perturbation for $|\psi_0^{(0)}\>$.
The first order perturbation energy is 
\beq
\varepsilon=\lambda^2\sum_{n=-\infty}^{\infty}v(\varepsilon+n\omega)J_n(f/\omega)^2.
\label{eq:large_w}
\eeq
In the discrete model, $v_d(E)$ is real for $|E|>2g$ with $E\in\mathbb{R}$.
Therefore, if $2g-\omega<\varepsilon<-2g$, $v_d(\varepsilon+n\omega)$ is real for any integer $n$, and Eq.~(\ref{eq:large_w}) has a real solution.
Now we assume $2g-\omega<\varepsilon<-2g$, which will be confirmed later.
Then for $n=0$, $v_d(\varepsilon)=-1/\sqrt{\varepsilon^2-4g^2}$ and $v_d(\varepsilon+n\omega)=O(1/n\omega)$ for $n\neq 0$.
In the leading order of $\omega$, therefore, we have $\varepsilon=\left[\lambda J_0(f/\omega)\right]^2v_d(\varepsilon)$.
By comparing it with the relation $E_d=\lambda^2v_d(E_d)$, we find that $\lambda$ is replaced by $\lambda J_0(f/\omega)$ and therefore $\varepsilon=E_d^{\rm eff}$.
There is a Floquet bound state and its quasi-energy is identical to that evaluated by the FM expansion.
Here we have considered only the case of $\omega\gg g$, but as long as $\omega\gtrsim 4g$, a Floquet bound state will exist because $\{\varepsilon+n\omega\}_{n\in\mathbb{Z}}$ can avoid the continuous spectrum $(-2g,2g)$, and hence the on-site potential $-n\omega+\lambda^2v(\varepsilon+n\omega)$ is real for any $n$.
On the other hand, for any $\omega<4g$, there exists an integer $n$ such that $\varepsilon+n\omega\in(-2g,2g)$, and hence the solution of Eq.~(\ref{eq:FEP_dot}) does not exist; there is no Floquet bound state, see Fig.~\ref{fig:hopping} (a).
The above result is consistent with the fact that the FM expansion is ensured to be convergent only for $\omega\gtrsim E_{\rm max}-E_{\rm min}\simeq 4g$.

Next let us consider the continuous model.
In this case, the continuous spectrum extends over $E>0$.
We therefore assume $\varepsilon<0$.
$v_c(\varepsilon)=-1/(2\sqrt{-g\varepsilon})$ and $v_c(\varepsilon+n\omega)=O(1/\sqrt{n\omega})$ for $n\neq 0$.
In the leading order of $\omega$, we have $\varepsilon\approx \left[\lambda J_0(f/\omega)\right]^2v_c(\varepsilon)$, which is solved as $\varepsilon\approx E_c^{\rm eff}$ (remember that $E_c^{\rm eff}$ is obtained by replacing $\lambda$ by $\lambda J_0(f/\omega)$ in $E_c$).
However, this is {\it not} a solution, because $\{\varepsilon+n\omega\}$ for $n>0$ are inevitably on the continuous spectrum, see Fig.~\ref{fig:hopping} (b), and we cannot neglect these contributions.
If we assume ${\rm Im}\varepsilon>0$, the imaginary part of $\varepsilon$ is evaluated by the perturbation theory as
\begin{align}
&{\rm Im}\varepsilon\approx\lambda^2\sum_{n=0}^{\infty}J_n\left(\frac{f}{\omega}\right)^2{\rm Im}v_c(\varepsilon+n\omega)
\nonumber \\
&\approx \lambda^2J_0\left(\frac{f}{\omega}\right)^2\left.\frac{dv_c(E)}{dE}\right|_{E={\rm Re}\varepsilon}{\rm Im}\varepsilon -\frac{\lambda^2}{\sqrt{g\omega}}\sum_{n=1}^{\infty}\frac{J_n(f/\omega)^2}{2\sqrt{n}}.
\end{align}
By substituting ${\rm Re}\varepsilon=E_c^{\rm eff}$, we obtain
\beq
{\rm Im}\varepsilon\approx -\frac{\lambda^2}{\sqrt{g\omega}}\frac{2}{4+\sqrt{3}}\sum_{n=1}^{\infty}\frac{J_n(f/\omega)^2}{\sqrt{n}} \equiv -\frac{\lambda^2}{\sqrt{g\omega}}A.
\eeq
It contradicts the assumption of ${\rm Im}\varepsilon>0$, and hence $\varepsilon_{\rm res}\equiv E_c^{\rm eff}-iA\lambda^2/\sqrt{g\omega}$ is not a solution of Eq.~(\ref{eq:FEP_dot}) and not a quasi-energy of the Floquet bound state.
However, this can be interpreted as the quasi-energy of the {\it Floquet resonant state} (see Ref.~\cite{Reed-Simon_text4} for mathematics of resonant states and Ref.~\cite{Hatano2008} for their physical meaning).
The imaginary part of $\varepsilon_{\rm res}$ corresponds to the inverse of the lifetime of this resonant state.
The real part of $\varepsilon_{\rm res}$ is identical to the quasi-energy of a localized state near the dot, which is obtained by the FM expansion.
Thus we conclude that {\it when the FM expansion is not convergent, the FM effective Hamiltonian does not give correct Floquet states, but its eigenstates represent the metastable states of the periodically driven system for large $\omega$}.
The description by $H_{\rm eff}^{(0)}$ is valid up to the lifetime of the metastable state.
The lifetime of the bound state of $H_{\rm eff}^{(0)}$ is $\tau_{\rm res}\sim\omega^{1/2}$ for large $\omega$ in our model~\endnote
{The behavior of ${\rm Im}\varepsilon\sim \omega^{-1/2}$ is {\it not} a universal feature.
It depends on $W(x)$ as to how ${\rm Im}\varepsilon$ depends on $\omega$ for large $\omega$.
In any case, however, ${\rm Im}\varepsilon\rightarrow 0$ in the limit of $\omega\rightarrow\infty$.}.

\begin{figure}[t]
\begin{center}
\includegraphics[clip,width=8cm]{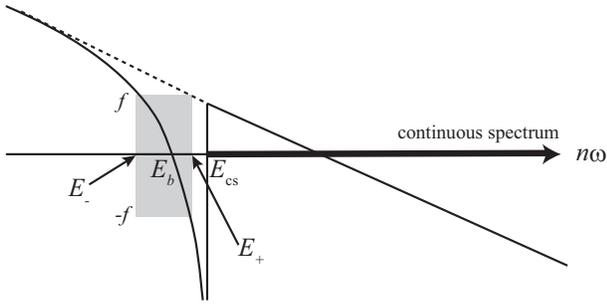}
\caption{A schematic picture of $U_n=-n\omega+\lambda^2{\rm Re}v(\varepsilon+n\omega)$ as a function of $n\omega$.
The minimum of the continuous spectrum is denoted by $E_{\rm cs}$, $E_{\rm cs}=-2g$ in the discrete model and $E_{\rm cs}=0$ in the continuous model.
The particle does the Bloch oscillation in the range $E_-\leq n\omega\leq E_+$, which is depicted as the shaded region.
}
\label{fig:potential}
\end{center}
\end{figure}

Starting from Eq.~(\ref{eq:FEP_dot}), we can also study the low-frequency regime, $\omega\ll |E_b|,g$.
The following analysis is applicable to both the discrete and continuous models.
A schematic picture of the on-site potential $-n\omega+\lambda^2{\rm Re}v(\varepsilon+n\omega)\equiv U_n$ is given in Fig.~\ref{fig:potential}.
First we neglect the imaginary part of $v(\varepsilon+n\omega)$.
We are interested in the eigenstate with the eigenvalue $\varepsilon\in[-\omega/2,\omega/2)$, and such an eigenstate will be localized near $n\omega\approx E_b$.
The eigenstate is expanded as $|\psi\>=\sum_{n=-\infty}^{\infty}\psi_n|n\>$.
Roughly speaking, the real wave function will behave as $\psi_n\approx {\rm Re}\left[B\exp\left(i\sum_{m=n^*}^nk_m\right)\right]\approx {\rm Re}\left[ B\exp\left(\frac{i}{\omega}\int_{E_b}^{n\omega}k(E)dE\right)\right]$, where $n^*$ is the integer closest to $E_b/\omega$, $k_m=k(m\omega)$ is the ``momentum'' determined by the ``energy conservation law'', $\varepsilon=f\cos k_m+U_m$, that is, $\cos k_m=(\varepsilon-U_m)/f\approx-U_m/f$, and $B\in\mathbb{C}$ is some constant.
The ``particle'' on the energy space oscillates around $n^*$ (the Bloch oscillation) in the range $E_-\leq n\omega\leq E_+$ satisfying $|U_n/f|\leq 1$, while the wave function decays exponentially outside of this region since $k_n$ becomes imaginary, see Fig.~\ref{fig:potential}.
The amplitude of the Bloch oscillation (the number of integers $n$ with $E_-\leq n\omega\leq E_+$) is roughly proportional to $f/\omega$, and hence $|B|^2\sim\omega/f$ due to the normalization of the wave function.
Because of the quantum tunneling in the energy space, there is a very small but finite amplitude for the particle to be found in the region of the continuous spectrum.
The imaginary part of $\varepsilon$ is evaluated as
\begin{align}
{\rm Im}\varepsilon\approx\lambda^2\sum_{\substack{n\in\mathbb{Z} \\ (n\omega\geq E_{\rm cs})}}{\rm Im}\left[v(\varepsilon+n\omega)\right]|\psi_n|^2
\nonumber \\
\sim\frac{\lambda^2}{\omega}\int_{E_{\rm cs}}^{\infty}dE|\psi(E)|^2{\rm Im}\lim_{\delta\rightarrow+0}v(E+i\delta),
\end{align}
where $\psi(n\omega)\equiv\psi_n$ and $E_{\rm cs}$ is the minimum value of the continuous spectrum.
Since $\psi(E)$ decays exponentially for $E_+<E<E_{\rm cs}$, 
\beq
|\psi(E)|^2\sim\frac{\omega}{f}e^{-\frac{E_{\rm cs}-E_+}{\omega\xi}}\left|e^{\frac{i}{\omega}\int_{E_{\rm cs}}^Ek(E')dE'}\right|^2
\eeq
for $E>E_{\rm cs}$, where $\xi>0$ is the decay length independent of $\omega$.
Therefore, remembering  ${\rm Im}\lim_{\delta\rightarrow+0}v(E+i\delta)<0$, we obtain
\begin{align}
{\rm Im}\varepsilon\sim -\frac{\lambda^2}{f}\exp\left[-\frac{E_{\rm cs}-E_+}{\omega\xi}\right],
\label{eq:low_w}
\end{align}
where we have ignored the nonessential constant factor.
Also in the low-frequency regime, there is no Floquet bound state, but the Floquet resonant state with the exponentially long lifetime $\tau_{\rm res}\sim\exp[(E_{\rm cs}-E_+)/\omega\xi]$ appears.
It means that, no matter how small $\omega$ is, the adiabatic theorem breaks down in extremely long timescale diverging exponentially in the limit of $\omega\rightarrow 0$.
In the previous study~\cite{Bukov2012}, the Luttinger model was investigated and it was shown that the lifetime of an intermediate state appearing in the low-frequency regime is relatively short, $\tau\sim 1/\omega$, which means that $\tau/T$ does not diverge as $\omega\rightarrow 0$.
The presence of the energy gap $E_{\rm cs}-E_b>E_{\rm cs}-E_+>0$ is crucial for the exponentially long lifetime which is understood as the quantum tunneling in the energy space.

More detailed analysis tells us that we cannot neglect ${\rm Im}v(\varepsilon+n\omega)$ for $E_-<E<E_+$, but the exponential dependence of the lifetime on $\omega$ is not altered by including this contribution, while the factor $\lambda^2/f$ in front of the exponential in Eq.~(\ref{eq:low_w}) is not reliable.
We do not go into more precise analysis in the low-frequency regime in this work.

In conclusion, we have investigated the Floquet eigenvalue problem of the periodically driven Friedrichs models and shown that, in the discrete model, the Floquet bound state exists and its quasi-energy is identical to that obtained by the FM effective Hamiltonian for sufficiently large $\omega$, while in the continuous model, exact Floquet states are quite different from eigenstates of the FM effective Hamiltonian.
However, in the latter case, there exists the Floquet resonant state and the real part of the resonant pole is identical to the energy of the bound state of the FM effective Hamiltonian.
The imaginary part of the resonant pole approaches zero as $\omega$ increases, which means that the bound state of the FM effective Hamiltonian is actually a metastable state whose lifetime is very long for large $\omega$.

We have studied specific one-particle models, but the result is instructive; even if the FM expansion is not convergent, the FM effective Hamiltonian can capture intermediate metastable states.
It is expected to be a general feature of periodically driven systems.

We have also studied the low-frequency regime and clarified that the quantum tunneling in the energy space induces nonadiabatic transitions, but its probability is exponentially small with respect to $\Delta E/\omega$, where $\Delta E$ is the energy gap between the continuous spectrum and the energy region of the ``Bloch oscillation'' around the bound state energy without driving.

We hope that this study triggers further studies such as metastability under periodic driving in many-body systems and the effect of dissipation for driven systems~\cite{Breuer2000,Kohn2001,Hone2009,Shirai2014b}.

The author thanks Tomotaka Kuwahara and Keiji Saito for stimulating discussions.
He acknowledges the JSPS Core-to-Core Program ``Non-equilibrium dynamics of soft matter and information''.

\end{document}